\documentclass[12pt]{article}

\usepackage{scicite}
\usepackage{times}
\usepackage{graphicx}
\usepackage{xcolor}
\usepackage{soul}

\newcommand{\rnew}[1]{\textcolor{red}{#1}}

\newcommand{\rep}[1]{representation}

\newenvironment{sciabstract}{%
\begin{quote} \bf}
{\end{quote}}

\title{Homological percolation transitions in growing simplicial complexes}

\author
{Y. Lee,$^{1}$ J. Lee,$^{1}$ S. M. Oh,$^{1}$ D. Lee,$^{1}$ B. Kahng$^{1\ast}$\\
\\
\normalsize{$^{1}$CCSS, CTP, Department of Physics and Astronomy, Seoul National University,}\\
\normalsize{Seoul 08826, Korea}\\
\\
\normalsize{$^\ast$To whom correspondence should be addressed; E-mail:  bkahng@snu.ac.kr}
}

\date{}

\begin{document} 

\baselineskip24pt

\maketitle 

\begin{sciabstract}
Simplicial complex (SC) representation is an elegant mathematical framework for representing the effect of complexes or groups with higher-order interactions in a variety of complex systems ranging from brain networks to social relationships. Here, we explore the homological percolation transitions (HPTs) of growing SCs using empirical datasets and a model proposed. The HPTs are determined by the first and second Betti numbers, which indicate the appearance of one- and two-dimensional macroscopic-scale homological cycles and cavities, respectively. A minimal SC model with two essential factors, namely, growth and preferential attachment, is proposed to model social coauthorship relationships. This model successfully reproduces the HPTs and determines the transition types as infinite order (the Berezinskii--Kosterlitz--Thouless type) with different critical exponents. In contrast to the Kahle localization observed in static random SCs, the first Betti number continues to increase even after the second Betti number appears. This delocalization is found to stem from the two aforementioned factors and arises when the merging rate of two-dimensional simplexes is less than the birth rate of isolated simplexes. Our results can provide topological insight into the maturing steps of complex networks such as social and biological networks. 
\end{sciabstract}

\section*{\rnew{Introduction}}

Complex systems are composed of elements that interact with each other in various unpredictable ways.
In the graph approach, interactions among elements of a complex system are represented by a graph comprising a set of vertices (or nodes) and a set of edges (or connections) between pairs of nodes (that denote the elements) \cite{Albert2002,Dorogovtsev2002,Newman2003,Boccaletti2006,Araujo2014,Lee2018}. This graph representation successfully explains such emerging phenomena as the formation of a giant component during percolation \cite{Cohen2000} and pandemics and epidemics \cite{Pastor-Satorras2015}. However, little attention has been paid to systems with high-order interactions, except for several representations that consider higher-order interactions.\cite{Arenas2008,Benson2016,Lambiotte2019,Battiston2020}. 
The hypergraph, which is a method for representing high-order interactions, is a generalization of a graph in which interactions involving more than two elements are represented by a hyperedge \cite{Berge1973}. The hypergraph is suitable for describing social networks that include groups, such as coauthorship networks \cite{Liu2015}, protein-interaction networks that include protein complexes \cite{Palla2005}, structural brain networks \cite{Sizemore2018}, and processes such as social contagion \cite{Jhun2019,DeArruda2020} and cooperative dynamics \cite{Burgio2020}. 
A simplicial complex (SC), which is composed of simplexes, is a particular form of hypergraph \cite{Aleksandrov1998}. A simplex is characterized by the dimension $d$, which implies that it is constituted by $d+1$ vertices. For instance, when $d=0$, 1, 2, and 3, the simplex is a point, line segment, triangle, and tetrahedron, respectively. In mathematical terms, a $d$-dimensional simplex (denoted as a $d$-simplex) is a convex hull of $d+1$ points, which is often described as filling the internal region of the simplex. Furthermore, the convex hull of any nonempty subset of a $d$-simplex is called the face of the $d$-simplex; each face is itself a simplex. This hierarchical structure enables the use of the elegant mathematical approach, namely, algebraic topology. Facets are a set of maximal faces of a given SC \cite{Munkres19}. The properties of this representation, such as the clustering coefficient \cite{Kartun-Giles2019} and types of phase transitions \cite{Costa2016,Linial2016}, differ from those of graph representations. Such differences can also be found in diverse systems and phenomena, for instance, activity-driven temporal models \cite{Petri2018}, ecosystems with interacting competitors \cite{Grilli2017}, social contagion models \cite{Iacopini2019}, random walks, and synchronization \cite{Schaub2020,Matamalas2020}.

Graphs are classified into two types: static graphs, wherein the number of $0$-simplexes is fixed; and growing graphs, wherein this number is increased with time. A percolation transition represents the formation of a giant cluster as the number of $1$-simplexes is increased. For instance, for the Erd\H{o}s--R\'enyi (ER) random graph model, $N$ vertices are present from the beginning, and the ER graph is therefore static. An edge connects a pair of vertices with probability $p$. The graph structure changes notably across a point $p_c=1/N$, beyond which a macroscopic-scale giant cluster emerges. This phenomenon is called a percolation transition, and $p_c$ is the percolation threshold. The density of nodes belonging to the giant cluster, $G(p)$, increases in a power-law manner as $G(p)\sim (p-p_c)^{\beta}$. Thus, the percolation transition is of second order. In contrast, for a random growing graph \cite{Callaway2001}, at each time step, a vertex is added to the system, and then an edge is connected with probability $p$ between two vertices selected randomly. A giant cluster is formed as $G(p)\sim \exp(-\alpha \sqrt{p-p_c})$; therefore, the percolation transition is of infinite order. Thus, the percolation transitions of static and growing graphs are of different types.
	 
Similarly, SCs are classified into static and growing types. The former maintains a fixed number of $0$-simplexes throughout the entire period of observation, whereas the latter has a $0$-simplex newly added to the system at each time step. As a generalization of the percolation transition, a homological percolation transition (HPT) was proposed  to reflect high-order interactions. For the static case, several models have been proposed, which extend the concept of an ER random graph to the SC. Particularly, an ER-like random SC model, which starts with fully connected $(d-1)$-simplexes and then creates $d$-simplexes with probability $p$, was introduced \cite{Linial2016}. The model exhibits an HPT wherein the density of the so-called shadow has a first-order transition. It was suggested that an HPT is related to the formation of macroscopic-scale giant loops or cavities, which are directly associated with the first or second Betti number \cite{Bobrowski2020}. Various other interesting features were also discovered \cite{Kahle2013,Costa2016}. However, an HPT for the growing case has not been investigated thus far.

\begin{figure}[!b]
\centering
\includegraphics[width=0.7\linewidth]{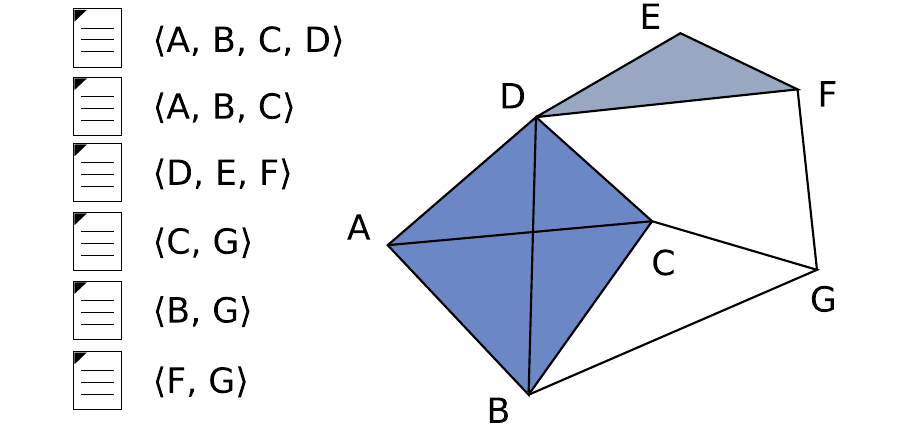}
\caption{\textbf{Schematic illustration of the SCR of coauthorship relationships}. Each list enclosed by angle brackets in the left panel represents a paper written by the listed authors. The SC in the right panel is the coauthorship complex constructed from these six papers. The facet degree of a vertex is the number of facets in which the vertex participates. For example, the facet degree of C is 2, and that of G is 3. B-C-G-B and B-D-C-G-B are examples of homologous cycles, as their symmetric difference, B-C-D-B, is the boundary of the $2$-simplex [B, C, D]. The B-A-D-C-G-B cycle is also homologous to them. They all represent the same voids. The cycles C-D-F-G and C-D-E-F-G-C are not homologous to them but are homologous to each other. Any cycle homologous to B-C-D-F-G-B can be represented as the symmetric difference of a cycle homologous to B-C-G-B and a cycle homologous to C-D-F-G up to a boundary cycle. The first Betti number is 2.}
\label{Fig1}
\end{figure}

We consider herein an HPT of growing SCs with empirical data and propose a minimal model. The empirical data are for coauthor relationship. We traced the papers and their authors that cite a few pioneering papers on a specific research field (network science) from the inception up to the end of 2017 for approximately 20 years. Each paper and its $n$ authors ($n=1,2,\cdots$) are represented herein by an  $(n-1)$-simplex and $n$ $0$-faces of the simplex, as illustrated in Fig.~\ref{Fig1} \cite{Newman2001,Fortunato2018,Kong2019}. As a new paper that cites any selected pioneering papers is published, this simplex is added to the existing SC so that the SC continues to grow. We investigated an HPT from this dataset. Moreover, a minimal model relevant to the HPT is proposed herein, and an investigation of the macroscopic properties is presented. To check whether the properties of HPT in growing simplcial complexes are universal, we have also considered a protein-interaction SC model including duplication, mutation, and divergence factors.

The evolution of coauthorship graphs of several research topics, including network science, were explored as a function of time from inception \cite{Lee2010}. The percolation transition was recognized as infinite order. Fig.~\ref{fig:n0_np_t}a ilustrates how the macroscopic properties of the coauthorship graph reach a mature state. The growth process is divided into three stages from the perspective of graph representation: 
(i) Small isolated components are created.
(ii) A treelike giant component is formed by merging clusters in the early stages, followed by the connection of long-range edges. Thus, long loops are formed in the later stage.
(iii) The network becomes entangled by forming intra-cluster edges. 
These three stages are characterized by the mean separation ($\bar d$) between two connected vertices averaged over different clusters. In stage (i), $\bar d$ remains almost constant; in stage (ii), $\bar d$ increases overall but with fluctuations; and in stage (iii), $\bar d$ decreases overall. These three stages are indicated in Fig.~\ref{fig:n0_np_t}b.  However, the classification based on $\bar d$ can be considered rather primitive owing to the lack of an appropriate mathematical tool in graph representation. In this paper, we demonstrate that the evolutionary steps can be reconstructed from the perspective of SC representation in terms of HPTs associated with the first and second Betti numbers. We elucidate portrays the route by which a coauthorship network reaches a mature state. Accordingly, the development and decline of a research topic through various intermediate processes can be discerned. 

\section*{\rnew{Results}}

\begin{figure}[!htb]
\centering
\includegraphics[width=\linewidth]{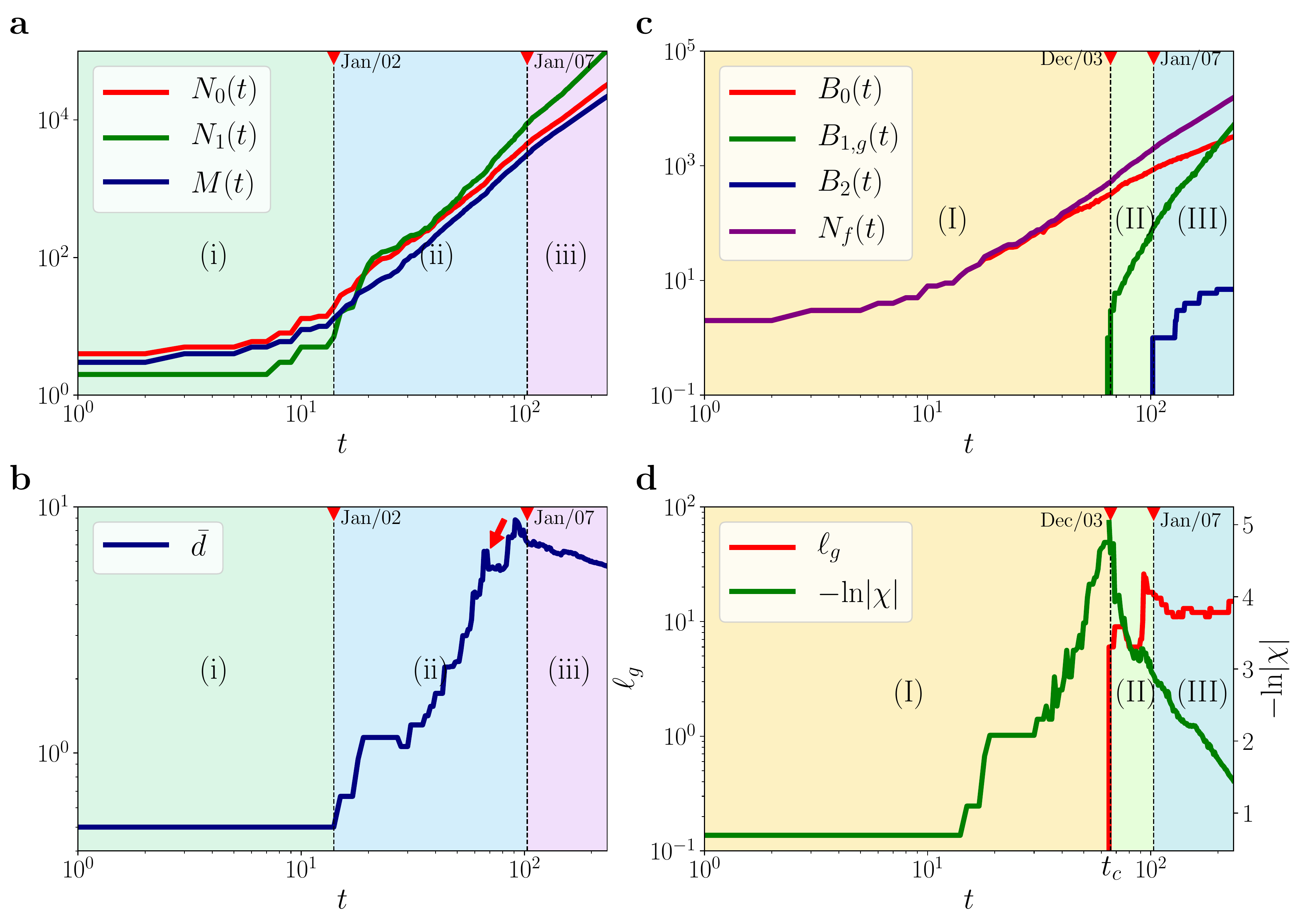} 
\caption{\textbf{Evolution of several graph and homological quantities.} (\textbf{a}) Plots of the number of vertices $N_0(t)$, the number of edges $N_1(t)$, and the size of the largest component $M(t)$ versus time step $t$ in graph representation. (\textbf{b}) Plots of the mean separation between two vertices $\bar d$ as a function of $t$ in the graph representation. (\textbf{c}) Plots of the zeroth Betti number $B_0$ (the number of components), the first Betti number $B_{1,g}$ (the number of homological one-dimensional cycles) of the largest cluster, the second Betti number $B_2$, and the number of facets $N_f$ as a function of time step $t$. (\textbf{d}) Plots of the length of the longest cycle $\ell_g$ and the logarithm of the Euler characteristic $-\ln |\chi|$ as a function of time step $t$.}
\label{fig:n0_np_t}
\end{figure}

\noindent \textbf{Homological percolation transitions.} The two HPTs occur successively \cite{Bobrowski2020} at transition points $t_{c1}$ and $t_{c2}$, as shown in Fig.~\ref{fig:n0_np_t}c. They are determined by the first and second Betti numbers in the giant cluster, denoted as $B_{1,g}$ and $B_{2,g}$, which represent the numbers of homological cycles and cavities in the giant cluster, respectively. The homological cycles also exist in finite clusters smaller than the giant cluster. Moreover, the giant cluster is also called an infinite cluster in percolation theory because the critical behavior of percolation transition is treated in the thermodynamic limit $N\to \infty$. Hence, the homological cycles in finite clusters can be ignored compared with those in the giant cluster. This is also confirmed by empirical data. In contrast, the second Betti number is nonzero only in the giant cluster. Thus, $B_{2,g}=B_2$.   

This homological classification scheme separates the regions of phases (i) and (ii) into new phases (I) and (II), as illustrated in Fig.~\ref{fig:n0_np_t}. In phase (I), both Betti numbers are zero; in phase (II), a giant one-dimensional homological cycle appears, that is, $B_{1,g} >0$, but $B_2$ is still zero. In phase (III), $B_2$ is also finite. The point $t_{c2}$ coincides with the previous point \cite{Lee2010}, which was determined intuitively. In addition, the homological properties of coauthorship complexes, such as simplicial contraction, facet size distribution, and persistent homology, have been explored in various coauthorship datasets \cite{Patania2017,Carstens2013}; however, these datasets were collected at specific times. Thus, the HPTs arising in evolutionary processes have not thus far been identified in real-world complexes.

\begin{figure}[!b]
\centering
\includegraphics[width=\linewidth]{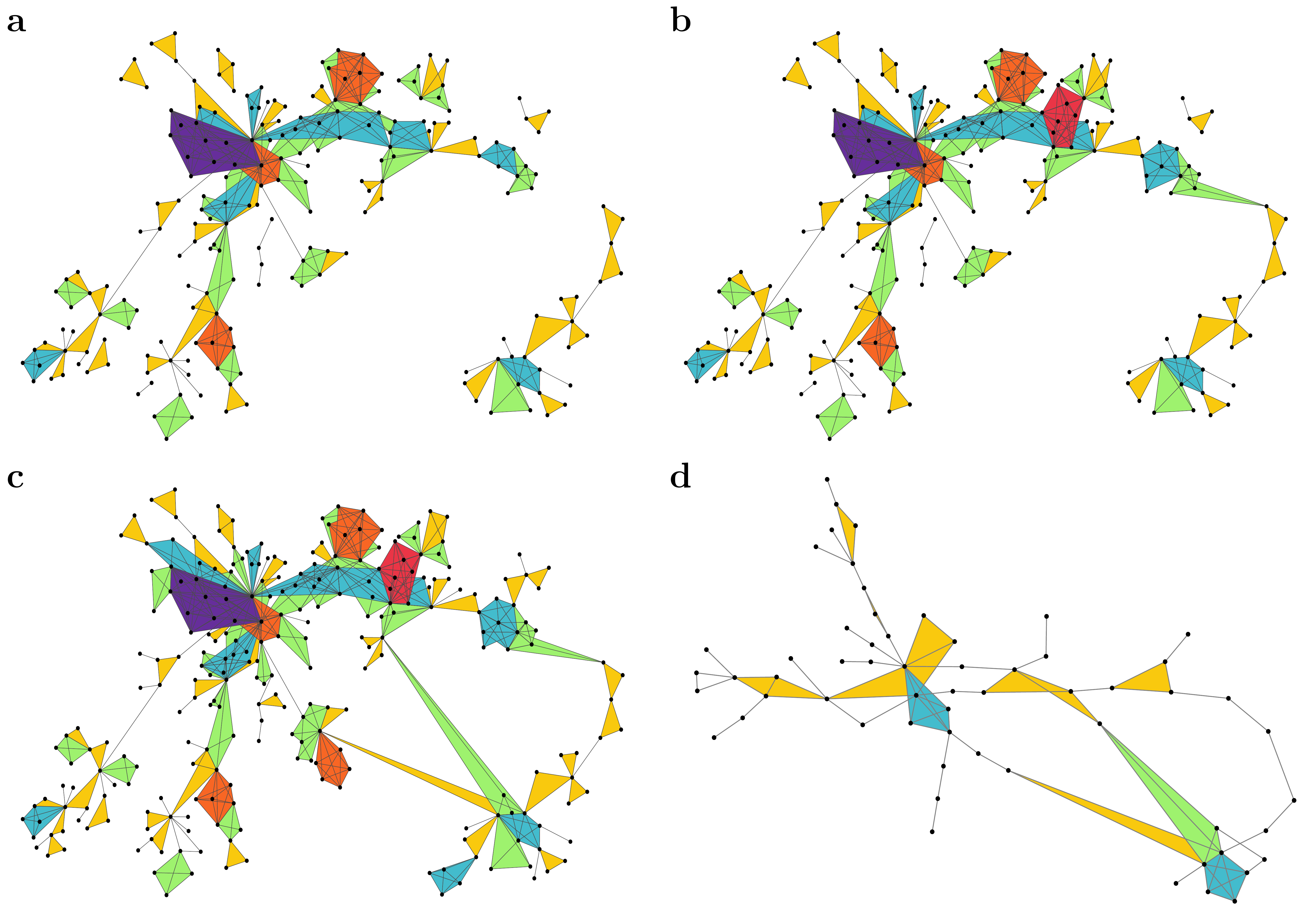}
\caption{\textbf{Snapshots of the coauthorship SC.} (\textbf{a}) Snapshot of the SC at $t=65$, where the second-largest SC has not yet merged with the giant SC. (\textbf{b}) Snapshot of the giant SC at $t=66$, where the second-largest component has merged. (\textbf{c}) Snapshot of the SC at $t=69$, where the long-range edges are connected. (\textbf{d}) Snapshot of SC after a process of homological simplification \cite{Wilkerson2013} of (\textbf{c}).}
\label{snapshot}
\end{figure}

Two important phenomena underlie the evolution of coauthorship networks: divergence and internal entanglement. When a student graduates from a university, she/he moves to a postdoctoral position in another group. This transfer enables both parties to broaden their experience and is thus beneficial to them. When the former student publishes a paper with her/his new colleagues, long-range connections are made between the old and new groups. These intergroup edges result in the formation of a long-distance homological loop. In particular, when the length of this loop is macroscopic, this giant loop results in an HPT in SC representation (see the snapshots in Fig.~\ref{snapshot}). In addition, the intragroup edges are also reinforced as the group members publish more papers together. This internal edge entanglement results in the formation of two-dimensional voids. Thus, another type of HPT occurs, in which the second Betti number becomes nonzero. The phenomena of divergence and internal entanglement correspond to the central factors in the evolution of biological networks, divergence, and mutation during reproduction \cite{Kim2002}.

Here, we specifically considered an HPT of the coauthorship complex $S(t)$. We traced the first Betti number of the largest cluster as a function of time step $t$, which is denoted as $B_{1,g}(t)$. We found that $B_{1,g}(t)$ first becomes nonzero at time step $t_{c1}=67$. It apparently exhibits a power-law increase as $B_{1,g}(t)-B_{1,g}(t_{c1}) \sim (t-t_{c1})^{2.1}$. We also measured the length of the longest homological cycle $\ell_g(t)$ as a function of $t$, as shown in Fig.~\ref{fig:n0_np_t}d. We find that $\ell_g$ suddenly increases at $t_{c1}$, at which the giant cluster acquires an interbranch edge, and a macroscopic-scale long cycle is formed, as shown in Fig.~\ref{snapshot}d. We regarded this point as the transition point of the first HPT. In retrospect, this point was identified in the graph representation, in terms of the mean separation $\bar d$, where $\bar d$ decreases noticeably, as indicated by the arrow in Fig.~\ref{fig:n0_np_t}b. It was reported previously that the length of a homological cycle is positively associated with the number of linked communities \cite{Patania2017}. Thus, the formation of a long cycle indicates the formation of global collaborations.   

We also identified the transition point $t_{c1}$ using the Euler characteristic \cite{Bobrowski2020}. It was proposed that the Euler characteristic of the giant cluster, which is defined as $\chi \equiv \sum_k (-1)^k B_{k,g}$, becomes zero near the transition point of the HPT. Here, we measured $\chi=B_{0,g}-B_{1,g}+B_{2,g}$, because we obtained $B_{k,g}=0$ for $k \ge 3$ and plot $-\ln |\chi|$ as a function of $t$. It was found to diverge at $t_{c1}$. 

We separated the region beyond $t_{c1}$ into two regimes, (II) and (III), using the second Betti number $B_2$. In regime (II), the first Betti number, $B_{1,g}(t)$, increases continuously with time. In contrast, $\ell_g$ increases abruptly and then decreases slowly and reaches a steady state in which $\ell_g$ is constant overall, with some fluctuations, whereas $\bar d$ decreases continuously. In this late regime, the SC becomes increasingly entangled as more papers are published within each group. We measured the second Betti number $B_2(t)$ to check for the formation of cavities enclosed by simplexes. There exists a nonzero second Betti number at a transition point $t_{c2}$, as shown in Fig.~\ref{fig:n0_np_t}c, beyond which $B_2(t)$ remains nonzero. We found that these cavities were formed in the giant cluster.

In Fig.~\ref{fig:n0_np_t}c, the first Betti number $B_{1,g}(t)$ increases continuously even after the second Betti number $B_2(t)$ appears and then increases. This behavior differs from that in the Kahle localization, wherein the first Betti number decays rapidly to zero for the static Erd\H{o}s--R\'enyi (ER)-type random complex model \cite{Kahle2013} as the second Betti number appears. This difference results from the fact that the coauthorship complex is growing, and isolated complexes are thus continuously generated and accumulate over time. Some of them merge with a giant complex and contribute to the formation of new homological cycles, and the first Betti number $B_1(t)$ increases. Therefore, the number of isolated clusters, $B_0(t)$, may be expected to decrease. However, this rate of decrease was lower than the rate of increase of $B_0(t)$ resulting from the creation of new complexes. Thus, both $B_0(t)$ and $B_1(t)$ increase. The first and second Betti numbers, $B_1(t)$ and $B_2(t)$, exhibit similar behaviors; they also increase together. This issue is discussed in detail in more subsequent sections. 

\begin{figure}[!h]
\centering\includegraphics[width=\linewidth]{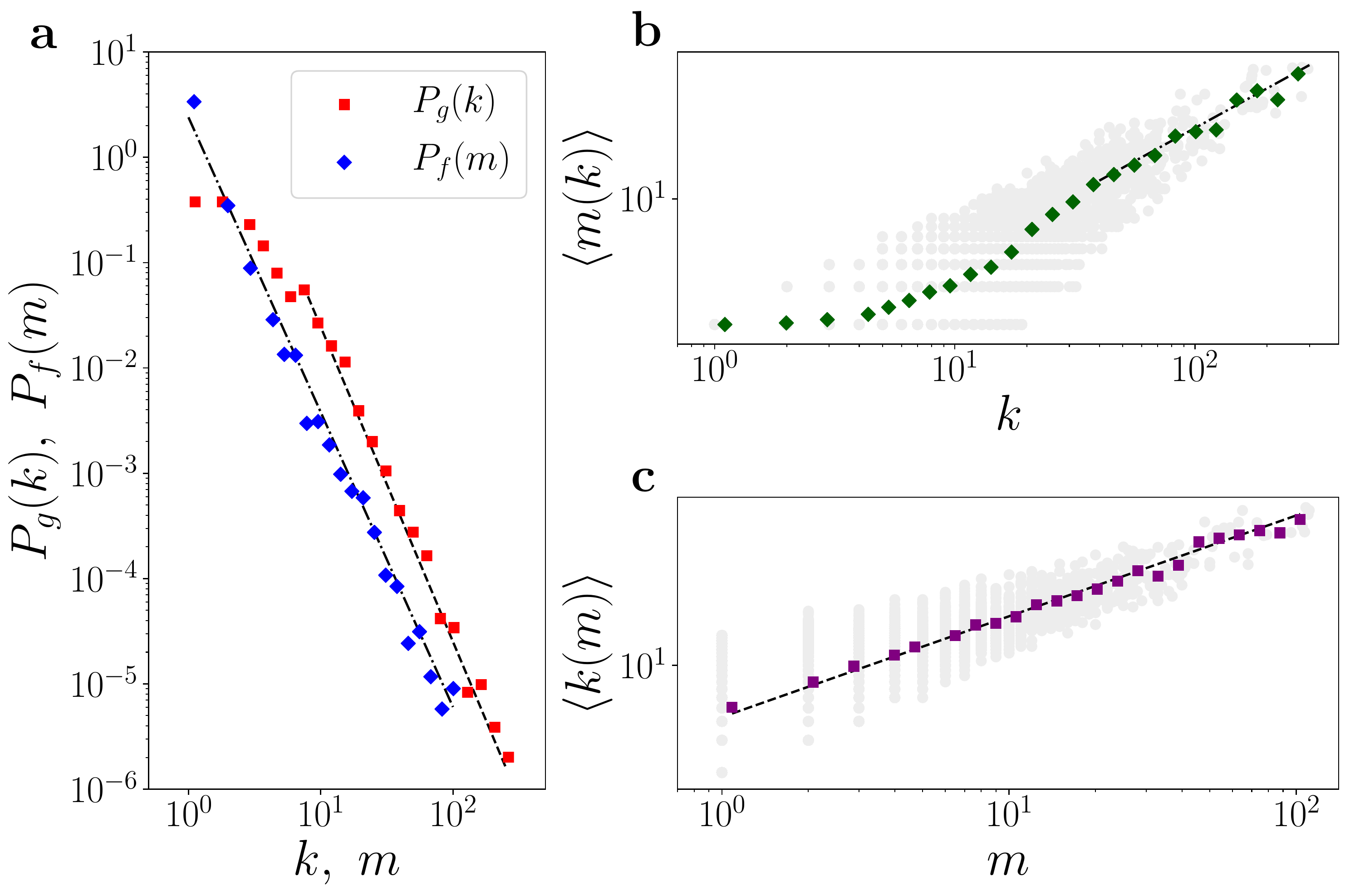} 
\caption{{\bf Graph and facet degree distributions.} (\textbf{a}) Plots of the graph degree distribution $P_{d,g}(k)$ and facet degree distribution $P_{d,f}(m)$ as functions of $k$ and $m$, respectively. (\textbf{b}) Plot of the mean facet degree $\langle m(k)\rangle$ of each vertex with graph degree $k$. The guide line (solid) has a slope of $1.05\pm 0.06$. (\textbf{c}) Plot of the mean graph degree $\langle k(m) \rangle$ of each vertex with facet degree $m$. The guide line (solid) has a slope of $0.91\pm 0.04$.}
\label{fig:degree_dist}
\end{figure}

\noindent{\bf Facet degree distribution.} In graph representation, the degree $k_i$ of vertex $i$ is the number of edges connected to vertex $i$. Here, this degree is referred to as the graph degree to distinguish it from the facet degree proposed below. We measured the graph degree of each vertex in the giant cluster and obtained the graph degree distribution, denoted as $P_{d,g}(k)$. The graph degree distribution exhibits power-law decay, $P_{d,g}(k)\sim k^{-\lambda_g}$, where $\lambda_g \approx 2.89\pm 0.06$. In SC representation, facets are the maximal faces of an SC. The facet degree $m_i$ of vertex $i$ is the number of facets to which the vertex $i$ belongs. The facet degrees in a giant cluster have a facet degree distribution $P_{d,f}(m)$, which is also called the simplicial degree distribution \cite{Patania2017}. We obtained $P_{d,f}(m)\sim m^{-\lambda_f}$, where $\lambda_f\approx 2.72\pm 0.11$, as shown in Fig.~\ref{fig:degree_dist}a. Thus, the exponents of the two degree distributions have slightly different values. 

To examine the correlation between the two degrees, we plotted the average facet degree $\langle m(k) \rangle$ of the vertices with graph degree $k$ in Fig.~4b and the average graph degree $\langle k (m)\rangle$ of the vertices with facet degree $m$ in Fig.~\ref{fig:degree_dist}. The vertices with a large (small) graph degree tend to have a large (small) facet degree, on average. However, fluctuations are unusual; when $k$ and $m$ are large, both the fluctuations of the facet degree for a given $k$ and those of the graph degree for a given $m$ are relatively small. Thus, the asymptotic behavior of the average quantities $\langle m(k) \rangle \sim k^{1.05\pm 0.06}$ and $\langle k(m) \rangle \sim m^{0.91\pm 0.04}$ are reciprocal. However, when $\langle k \rangle$ and $\langle m \rangle$ are small, the fluctuations of both are relatively large. This is because the field of network science includes interdisciplinary research subjects, such as mathematics, theoretical physics, and biology, where the number of authors per paper varies widely from one to more than 10 people. Moreover, a few review papers have large graph degrees but small facet degrees. In contrast, when the dimensions of the complexes are homogeneous, the two degree distributions have the same degree exponents, as we show later for a simple model.

\noindent{\bf Minimal model.} We propose a minimal model of the HPT that occurs in a growing complex. At each time step, a new vertex is added to the system; then, three vertices are selected with probability proportional to $m_i+a$, where $m_i$ is the facet degree of node $i$, and $a$ is a constant. The vertices are connected with probability $p$. This triangle is regarded as a two-dimensional simplex in the SCR. This process is repeated $t$ times. This model is an extension of the previous model of a randomly growing graph \cite{Callaway2001} in which two randomly selected vertices are connected with probability $p$. It differs from the previous model in that a $2$-simplex rather than a $1$-simplex is added, and the three nodes are selected according to their facet degrees rather than randomly. 

This growing SC model exhibits an infinite-order percolation transition in graph representation at a transition point $p_{c0}$. The cluster size distribution for $p < p_{c0}$ exhibits power-law decay, where the exponent depends on $p$ and $a$. However, the cluster size distribution of finite clusters for $p > p_{c0}$ decays exponentially. The transition point approaches zero as $a\to 0$. The giant cluster size $G(p,t)$ increases significantly with time as $G(p,t)\approx G(p)t$ asymptotically. In the steady state, $G(p)$ exhibits an essential singular behavior: $G(p)\sim \exp[-\alpha_0(p-p_{c0})^{-\beta_0}]$, where $\alpha_0$ is a nonuniversal constant. The transition point, $p_{c0}\approx 0.0031$, and exponent, $\beta_0\approx 0.44$, are obtained for $a=0.1$. For comparison, the exponent $\beta=1/2$ for the growing random network model~\cite{Callaway2001}. We present the analytic solutions of the percolation transition in this minimal model for general $a$ in the SI Appendix. 

We considered the graph degree distribution $P_{d,g}(k)$ and analytically obtain $P_{d,g}(k)\sim (k+a)^{-\lambda_g}$, where $\lambda_g=2+a/(3p)$. However, a pair of $2$-simplexes are more likely to be connected by sharing a vertex than by sharing an edge in large systems because the first case occurs with probability $O(1/N)$, whereas the second case occurs with probability $O(1/N^2)$. Hence, the graph and facet degrees of each vertex depend linearly on each other. The facet degree distribution exhibits power-law decay with the same exponent value, that is, $\lambda_f=\lambda_g$, as $P_f(m)\sim P_g(k)$. We confirmed this result using numerical simulations. A detailed derivation is presented in the SI Appendix.      

We counted the first Betti number, that is, the number of homological cycles, numerically as a function of $t$; it shows extensive behavior: $B_1(p)\approx b_1(p)t$ asymptotically. The first Betti number in the steady state, $b_1(p)$, exhibits a transition. The transition point $p_{c1}$ is consistent with that for the percolation transition $p_{c0}$ such that $b_1(p)$ is zero for $p \le p_{c1}$ and finite for $p > p_{c1}$. Simulations and finite-size scaling analysis revealed that the first Betti number $b_1$ increases in a similar way such that $b_1(p)\sim \exp[-\alpha_1(p-p_{c1})^{-\beta_1}]$, where $\alpha_1$ is a nonuniversal constant, and the exponent $\beta_1\approx 0.44$ is the same as that for the percolation transition for $a=0.1$. This result may reflect the property of ER random networks that a giant cluster and long-range loop emerge at the same transition point in the thermodynamic limit. Thus, $b_1(p)$ behaves similarly to $G(p)$ near the transition point. 

The second Betti number $B_2(t)$ behaves nonextensively with respect to the number of $0$-simplexes $t$, but is proportional to $t^{0.7}$ asymptotically. Thus, $B_2(t)$ is written as $B_2(t)\approx b_2(p)t^{0.70}$ asymptotically. Simulations and finite-size scaling analysis show that $p_{c2}\approx 0.053(3)$ for $a=0.1$. This transition point differs from $p_{c1}\approx 0.0031$ for $B_1(p)$ for the same value, $a=0.1$. $b_2(p)$ also exhibits the essential singular form $b_2(p)\sim \exp[-\alpha_2(p-p_{c2})^{-\beta_2}]$, where $\alpha_2$ is a nonuniversal constant, and $\beta_2\approx 0.99$ for $a=0.1$. The transition points and exponent values for the other values of $a$ are listed in the SI Appendix.

\begin{figure}[]
\centering
\includegraphics[width=0.9\linewidth]{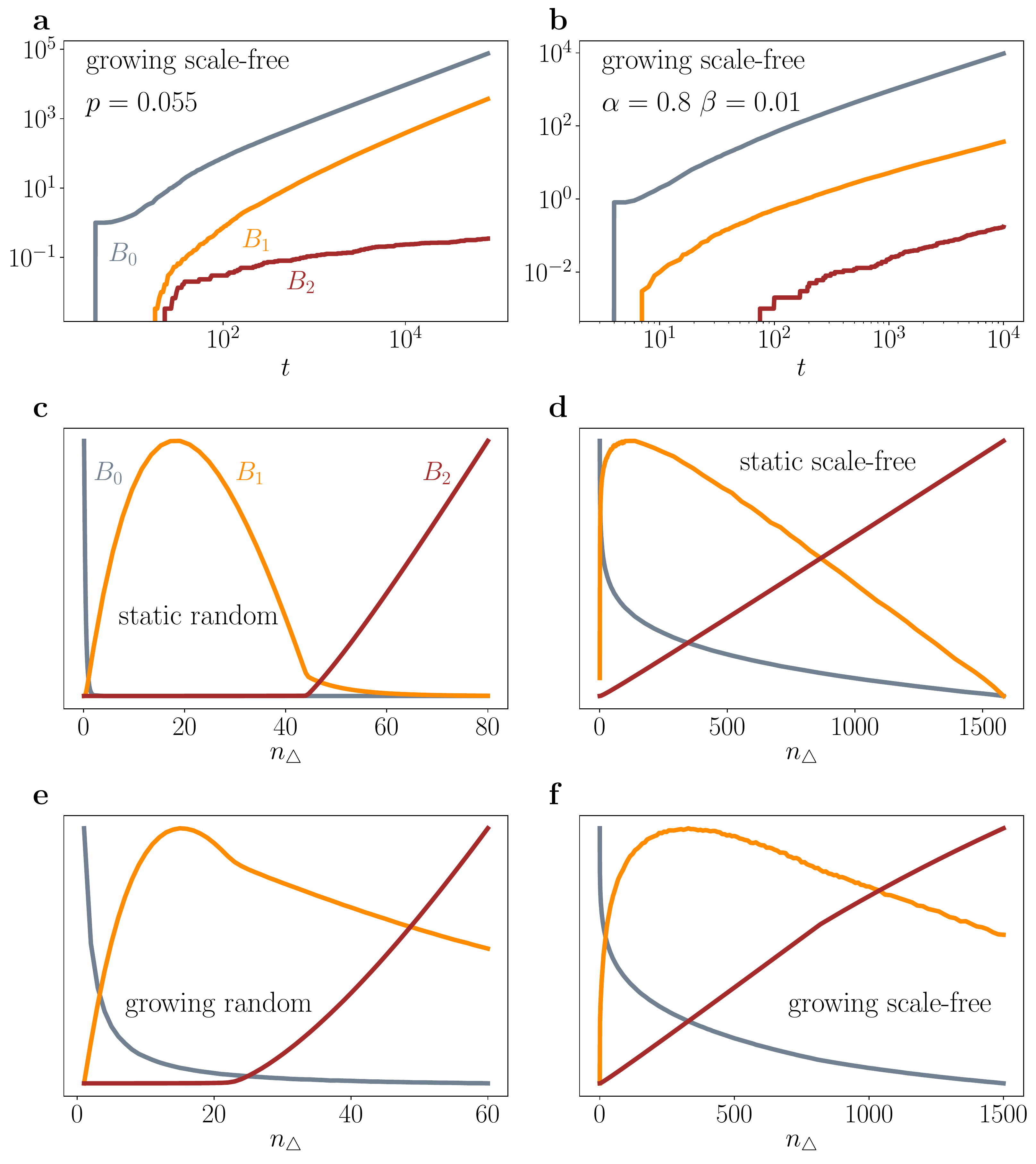} 
\caption{\textbf{Comparison of the first three Betti numbers of a static random SC and growing scale-free SC.} Plot of the Betti numbers versus time $t$ for (\textbf{a}) the minimal model of growing scale-free SC at $p = 0.055$, just above $p_{c2}\approx 0.053$ and for (\textbf{b}) the protein-interaction SC model at the model parameter values $\alpha=0.8$ and $\beta=0.01$.
(\textbf{c}) Plot of the Betti numbers versus the number of triangles per system size for the static random SC model with system size $N=100$. 
For easier viewing, we adjusted each Betti number (the scale of the vertical axis) appropriately.
(\textbf{d}) Similar to (c), but for the static scale-free SC model. 
(\textbf{e} and \textbf{f}) Similar to (c) and (d), but for the growing SC model. For (d) and (f), the localization of the Betti numbers does not occur. In other words, for (c) and (e), the second Betti number begins to increase only after the first Betti number reaches its maximum. }
\label{fig:kahle}
\end{figure} 

\noindent{\bf Kahle localization.} We reconsidered the localization of the Betti numbers in the minimal model \cite{Kahle2013}. To reproduce the evolution of the Betti numbers shown in Fig.~\ref{fig:n0_np_t}c, we numerically simulated the minimal model with a fixed $p >  p_{c2}$ because the second Betti number is not generated otherwise. The three Betti numbers $B_0(t)$, $B_1(t)$, and $B_2(t)$ were obtained as a function of $t$. As shown in Fig.~\ref{fig:kahle}a, the three Betti numbers appear successively, and they all increase with time. The first and second Betti numbers behave similarly to those we obtained from the coauthorship complex dataset. The Kahle localization does not occur in the minimal growing model.

To investigate the key factor affecting the localization, we considered a static model in which $N=100$ nodes exist continuously from the beginning. At each time step, three nodes were selected randomly and connected by a $2$-simplex. This process was repeated $N_{\triangle}=n_{\triangle}N$ times. Thus, the parameter $p$ is absent. The three Betti numbers are calculated as a function of $n_\triangle$ and are presented in Fig.~\ref{fig:kahle}b. We confirmed that the Kahle localization indeed occurs. Next, the static model was modified so that three nodes are selected with a probability proportional to their facet degree $m_i$ ($i=1,2,3$) in the form of $m_i+a$. As shown in Fig.~\ref{fig:kahle}c, the first Betti number immediately increases dramatically and then decreases slowly, whereas the second Betti number increases slowly. Therefore, the first and second Betti numbers coexist for a long time. After the first Betti number vanishes, the second Betti number continues to increase. Next, we considered the case of a growing complex. Initially, we set $N_{m0}=25$ nodes. At each time step, a node is added to the system, and three nodes are selected randomly (Fig.~\ref{fig:kahle}d) or according to the degree-dependent rule in Fig.~\ref{fig:kahle}e. In Fig.~\ref{fig:kahle}d, there exists a finite transition point for the second Betti number; however, the first Betti number increases immediately while remaining finite but decreases thereafter. At the transition point of the second Betti number, the decreasing rate of $B_1(t)$ is changed. In Fig.~\ref{fig:kahle}e, the first and second Betti numbers exhibit behavior similar to that shown in Fig.~\ref{fig:kahle}c. This result demonstrates that the localization behavior occurs only for the static random case in Fig.~\ref{fig:kahle}b.

\section*{\rnew{Discussion}}

We investigated the HPTs of growing SCs using the empirical data of coauthorship simplicial complexes and a model study. HPTs were identified by the Betti numbers. We revealed that the first three Betti numbers $B_k$ ($k=0,1$ and 2) are nonzero and the others are zero in the empirical dataset. This implies that papers with three authors play important role in the formation of SCs. Thus, we propose a minimal model composed of $2$-simplexes. The model was designed to include the growth and preferential attachment rules, which commonly appear in various complex systems. Owing to these factors, the first Betti number continues to increase in time even after the second Betti number appears. This implies that the coauthorship SCs are still developing, at least above $p_{c2}$ (see Fig.~\ref{fig:n0_np_t}c and Fig.~\ref{fig:kahle}a). 
To check whether the behaviors of three Betti numbers are universal or model-dependent, we modified an existing protein-interaction network model in graph representation \cite{Kim2002} into an SC version. The detailed rules are presented in the SI Appendix. We find that the first three Betti numbers behave similarly to the previous ones (see Fig.~\ref{fig:kahle}b). The pattern of the three Betti numbers as a function of $t$ is similar to that of the random growing SF model but differs from the localization pattern of Betti numbers for the static ER-like random SC model \cite{Kahle2013} (see Fig.~\ref{fig:kahle}c). For the localization and delocalization of the Betti numbers as a function of the number of triangles $n_{\triangle}$, we uncovered more detailed properties using the minimal model (see Fig. ~\ref{fig:kahle}c-~\ref{fig:kahle}f). 
Finally, we revealed that the HPTs in growing SCs are of infinite order, which is an intrinsic characteristic of growing graphs \cite{Callaway2001} and SCs.

The formation of a long-range cycle or loop is a significant factor for understanding the properties of diverse problems in physical and complex systems, for instance, phase transitions in equilibrium and nonequilibrium physical systems, and information spread in complex systems. Mean-field solutions for phase transitions of percolation and spin models are equivalent to the solutions on the Bethe lattice (tree structure) \cite{Dorogovtsev2008}. However, in lower dimensions, the mean-field solution is not correct because the effect of loop structure is significant. Thus, it is necessary to estimate when loop is formed, what loop size scales to system size, etc. For the ER model in graph representation, it was revealed that a macroscopic-scale loop is formed at a percolation threshold and the loop size is scaled as $\sim N^{1/3}$ with system size $N$ \cite{Bollobas2001}. Using this scaling, one can estimate the so-called golden time \cite{Lee2017,Choi2018} and others. In SC representation, however, such important and challenging problems have not been solved thus far, even though there exists an elegant mathematical tools such as algebraic topology. This study was intended as a first step toward such a novel approach.

\section*{\rnew{Empirical dataset}}
The main dataset used in this study is composed of papers on network science that cite two pioneering papers, those on the Watts--Strogatz model of small-world networks and the Barabasi--Albert model of scale-free networks \cite{Watts1998,Barabasi1999}, and highly cited early stage review papers \cite{Albert2002,Dorogovtsev2002,Newman2003,Boccaletti2006}. The dataset contains 21,653 papers authored by 32,016 distinct researchers and published from the beginning of the field in June 1998 through the end of 2017. The time $t$ is counted in units of months starting in June 1998, the month in which the small-world paper was published. At each time step, we constructed the coauthorship graph ${G}(t)$. In ${G}(t)$, the number of vertices (distinct authors) $N_0(t)$ and edges $N_1(t)$ increase with time $t$ as $N_0(t) \sim t^{2.6}$ and $N_1(t)\sim t^{3.2}$ asymptotically, as shown in Fig.~\ref{fig:n0_np_t}a. The months and $N_0$ values as a function of time step $t$ are listed in the SI Appendix. 

\bibliography{main}

\begin{thebibliography}{10}

\bibitem{Albert2002}
R.~Albert, A.-L. Barab{\'{a}}si, {\it Reviews of Modern Physics\/} {\bf 74}, 47
  (2002).

\bibitem{Dorogovtsev2002}
S.~N. Dorogovtsev, J.~F.~F. Mendes, {\it Advances in Physics\/} {\bf 51}, 1079
  (2002).

\bibitem{Newman2003}
M.~E.~J. Newman, {\it SIAM Review\/} {\bf 45}, 167 (2003).

\bibitem{Boccaletti2006}
S.~Boccaletti, V.~Latora, Y.~Moreno, M.~Chavez, D.~Hwang, {\it Physics
  Reports\/} {\bf 424}, 175 (2006).

\bibitem{Araujo2014}
N.~Ara{\'{u}}jo, P.~Grassberger, B.~Kahng, K.~J. Schrenk, R.~M. Ziff, {\it The
  European Physical Journal Special Topics\/} {\bf 223}, 2307 (2014).

\bibitem{Lee2018}
D.~Lee, B.~Kahng, Y.~S. Cho, K.-I. Goh, D.-S. Lee, {\it Journal of the Korean
  Physical Society\/} {\bf 73}, 152 (2018).

\bibitem{Cohen2000}
R.~Cohen, K.~Erez, D.~Ben-Avraham, S.~Havlin, {\it Physical Review Letters\/}
  {\bf 85}, 4626 (2000).

\bibitem{Pastor-Satorras2015}
R.~Pastor-Satorras, C.~Castellano, P.~{Van Mieghem}, A.~Vespignani, {\it
  Reviews of Modern Physics\/} {\bf 87}, 925 (2015).

\bibitem{Arenas2008}
A.~Arenas, A.~Fern{\'{a}}ndez, S.~Fortunato, S.~G{\'{o}}mez, {\it Journal of
  Physics A: Mathematical and Theoretical\/} {\bf 41}, 224001 (2008).

\bibitem{Benson2016}
A.~R. Benson, D.~F. Gleich, J.~Leskovec, {\it Science\/} {\bf 353}, 163 LP
  (2016).

\bibitem{Lambiotte2019}
R.~Lambiotte, M.~Rosvall, I.~Scholtes, {\it Nature Physics\/} {\bf 15}, 313
  (2019).

\bibitem{Battiston2020}
F.~Battiston, {\it et~al.\/}, {\it Networks beyond pairwise interactions:
  Structure and dynamics\/} {\bf 874}, 1 (2020).

\bibitem{Berge1973}
C.~Berge, E.~Minieka, {\it {Graphs and Hypergraphs}\/}, Graphs and Hypergraphs
  (North-Holland Publishing Company, 1973).

\bibitem{Liu2015}
P.~Liu, H.~Xia, {\it Scientometrics\/} {\bf 103}, 101 (2015).

\bibitem{Palla2005}
G.~Palla, I.~Der{\'{e}}nyi, I.~Farkas, T.~Vicsek, {\it Nature\/} {\bf 435}, 814
  (2005).

\bibitem{Sizemore2018}
A.~E. Sizemore, {\it et~al.\/}, {\it Journal of Computational Neuroscience\/}
  {\bf 44}, 115 (2018).

\bibitem{Jhun2019}
B.~Jhun, M.~Jo, B.~Kahng, {\it Journal of Statistical Mechanics: Theory and
  Experiment\/} {\bf 2019}, 123207 (2019).

\bibitem{DeArruda2020}
G.~F. de~Arruda, G.~Petri, Y.~Moreno, {\it Physical Review Research\/} {\bf 2},
  23032 (2020).

\bibitem{Burgio2020}
G.~Burgio, J.~T. Matamalas, S.~G{\'{o}}mez, A.~Arenas, {\it Entropy\/} {\bf
  22}, 744 (2020).

\bibitem{Aleksandrov1998}
P.~S. C. N. Q. S.~. Aleksandrov, {\it {Combinatorial topology}\/} (Dover
  Publications, Mineola, N.Y, 1998).

\bibitem{Munkres19}
J.~R. Munkres, {\it {Elements of algebraic topology}\/} (Perseus Books,
  Cambridge, Mass, 19), 12th edn.

\bibitem{Kartun-Giles2019}
A.~P. Kartun-Giles, G.~Bianconi, {\it Chaos, Solitons {\&} Fractals: X\/} {\bf
  1}, 100004 (2019).

\bibitem{Costa2016}
A.~Costa, M.~Farber, {\it Configuration Spaces\/}, F.~Callegaro, {\it
  et~al.\/}, eds. (Springer International Publishing, Cham, 2016), vol.~14, pp.
  129--153.

\bibitem{Linial2016}
N.~Linial, Y.~Peled, {\it Annals of Mathematics\/} {\bf 184}, 745 (2016).

\bibitem{Petri2018}
G.~Petri, A.~Barrat, {\it Physical Review Letters\/} {\bf 121}, 228301 (2018).

\bibitem{Grilli2017}
J.~Grilli, G.~Barab{\'{a}}s, M.~J. Michalska-Smith, S.~Allesina, {\it Nature\/}
  {\bf 548}, 210 (2017).

\bibitem{Iacopini2019}
I.~Iacopini, G.~Petri, A.~Barrat, V.~Latora, {\it Nature Communications\/} {\bf
  10}, 2485 (2019).

\bibitem{Schaub2020}
M.~T. Schaub, A.~R. Benson, P.~Horn, G.~Lippner, A.~Jadbabaie, {\it SIAM
  Review\/} {\bf 62}, 353 (2020).

\bibitem{Matamalas2020}
J.~T. Matamalas, S.~G{\'{o}}mez, A.~Arenas, {\it Physical Review Research\/}
  {\bf 2}, 12049 (2020).

\bibitem{Callaway2001}
D.~S. Callaway, J.~E. Hopcroft, J.~M. Kleinberg, M.~E.~J. Newman, S.~H.
  Strogatz, {\it Physical Review E\/} {\bf 64}, 41902 (2001).

\bibitem{Bobrowski2020}
O.~Bobrowski, P.~Skraba, {\it Physical Review E\/} {\bf 101}, 32304 (2020).

\bibitem{Kahle2013}
M.~Kahle, {\it arXiv:1301.7165 [math]\/}  (2013).

\bibitem{Newman2001}
M.~E.~J. Newman, {\it Proceedings of the National Academy of Sciences\/} {\bf
  98}, 404 (2001).

\bibitem{Fortunato2018}
S.~Fortunato, {\it et~al.\/}, {\it Science\/} {\bf 359}, eaao0185 (2018).

\bibitem{Kong2019}
X.~Kong, Y.~Shi, S.~Yu, J.~Liu, F.~Xia, {\it Journal of Network and Computer
  Applications\/} {\bf 132}, 86 (2019).

\bibitem{Lee2010}
D.~Lee, K.-I. Goh, B.~Kahng, D.~Kim, {\it Physical Review E\/} {\bf 82}, 26112
  (2010).

\bibitem{Patania2017}
A.~Patania, G.~Petri, F.~Vaccarino, {\it EPJ Data Science\/} {\bf 6}, 18
  (2017).

\bibitem{Carstens2013}
C.~J. Carstens, K.~J. Horadam, {\it Mathematical Problems in Engineering\/}
  {\bf 2013}, 1 (2013).

\bibitem{Kim2002}
J.~Kim, P.~L. Krapivsky, B.~Kahng, S.~Redner, {\it Physical Review E\/} {\bf
  66}, 55101 (2002).

\bibitem{Wilkerson2013}
A.~C. Wilkerson, T.~J. Moore, A.~Swami, H.~Krim, {\it ICASSP 2013 - 2013 IEEE
  International Conference on Acoustics, Speech and Signal Processing
  (ICASSP)\/} (IEEE, 2013), pp. 5258--5262.

\bibitem{Dorogovtsev2008}
S.~N. Dorogovtsev, A.~V. Goltsev, J.~F.~F. Mendes, {\it Reviews of Modern
  Physics\/} {\bf 80}, 1275 (2008).

\bibitem{Bollobas2001}
B.~Bollob{\'{a}}s, {\it et~al.\/}, {\it {Random Graphs}\/}, Cambridge Studies
  in Advanced Mathematics (Cambridge University Press, 2001).

\bibitem{Lee2017}
D.~Lee, W.~Choi, J.~Kert{\'{e}}sz, B.~Kahng, {\it Scientific Reports\/} {\bf
  7}, 5723 (2017).

\bibitem{Choi2018}
W.~Choi, D.~Lee, J.~Kert{\'{e}}sz, B.~Kahng, {\it Physical Review E\/} {\bf
  98}, 12311 (2018).

\bibitem{Watts1998}
D.~J. Watts, S.~H. Strogatz, {\it Nature\/} {\bf 393}, 440 (1998).

\bibitem{Barabasi1999}
A.-L. Barab{\'{a}}si, R.~Albert, {\it Science\/} {\bf 286}, 509 (1999).

\end{thebibliography}

\bibliographystyle{Science}

\section*{Acknowledgments}
{\bf Funding}: NRF, Grant No.~NRF-2014R1A3A2069005 (BK). \\
{\bf Competing interests}: The authors declare that they have no competing interests. \\ 
{\bf Data and materials availability}: All data used in this work are available from the authors upon reasonable request.

\end{document}


\newlength{\myfigwidth}
\setlength{\myfigwidth}{8cm}
		
\title{Supplemental Materials for \\ Homological percolation transitions in growing simplicial complexes}
\author{Yongsun Lee}
\affiliation{CCSS, CTP and Department of Physics and Astronomy, Seoul National University, Seoul 08826, Korea}
\author{Jongshin Lee}
\affiliation{CCSS, CTP and Department of Physics and Astronomy, Seoul National University, Seoul 08826, Korea}
\author{S. M. Oh}
\affiliation{CCSS, CTP and Department of Physics and Astronomy, Seoul National University, Seoul 08826, Korea}
\author{Deokjae Lee}
\affiliation{CCSS, CTP and Department of Physics and Astronomy, Seoul National University, Seoul 08826, Korea}
\author{B. Kahng}
\email{bkahng@snu.ac.kr}
\affiliation{CCSS, CTP and Department of Physics and Astronomy, Seoul National University, Seoul 08826, Korea}
\email{bkahng@snu.ac.kr}
\date{\today}
\nopagebreak
\maketitle
\nopagebreak
	
In this Supplemental Materials, we introduce details of the mimimal model for the homological percolation transitions (HPT) and present analytic solutions of percolation transitions in S1. Numerical simulation method and results for HPTs are also presented in S1. Next, we present the relation between the number of nodes and dates in S2.  
	
\section{The minimal model for growing scale-free complexes}
\subsection{Model}

The dynamics starts with $N_{0} \ge 3$ isolated nodes. At each time step, a new node is added to the system and then three nodes are selected with the probabilities proportional to $k_i+a$ for each node $i=1,2,$ and 3, where $k_i$ is the facet degree of node $i$ and $a$ is constant. The three nodes are connected and regarded as a 2-simplex, which is added to the system with probability $p$. These processes are repeated until $t$ times. Then the total number of nodes at time step $t$ in the system is $N(t)=N_0+t$ and the average number of facets is $pt$. The system consists of clusters of different sizes as time goes on. 

\subsection{The cluster-size distribution, giant cluster, and mean cluster size}

First, we are interested in the cluster size distribution. For analytical feasibility, we assume that all clusters are formed in the star-triangle shape that all triangles (facets) of a cluster are connected by sharing one node. Then the total facet degrees of nodes belonging to the cluster of size $s$ is written as $\frac{3}{2}(s-1)$. Then the probability that any one of nodes in a cluster of size $s_{i}$ is selected is proportional to $\Pi_{i=1}^{3} \Bigl( \frac{3}{2}(s_{i}-1)+ as_{i} \Bigr)$, where $a$ is the initial attractiveness of each node. Thus the rate equation of $N_s(p,t)$, defined as the number of cluster of size $s$, becomes
\begin{align}
\label{eq:rate_eq_of_n_s_p_t_for_scale_free_with_facet_degree_for_d_3}
\frac{d (N_s(p,t))}{d t} &= p\Bigl[ \sum_{i,j,k=1}^{\infty}{ ( J_{i}N_{i} J_{j}N_{j} J_{k}N_{k} ) \delta_{i+j+k, s}} + { \sum_{ i, k=1 }^{\infty} { 3 J_{i}N_{i} \Bigl(\frac{i}{N} \Bigr) ( J_{k}N_{k} ) \delta_{i + k, s}} }-3J_{s}N_s - {3 J_{s}N_s \Bigl(\frac{s}{N} \Bigr)} \Bigr] +\delta_{1s},
\end{align}
where $J_{s}(t) \equiv (s(3/2+a)-3/2)(3pt+aN(t))^{-1}$. In the limit $t\to \infty$, the second and fourth terms in the R.H.S. of Eq.~\eqref{eq:rate_eq_of_n_s_p_t_for_scale_free_with_facet_degree_for_d_3} become negligible. Eq.~\eqref{eq:rate_eq_of_n_s_p_t_for_scale_free_with_facet_degree_for_d_3} is then reduced to 
\begin{align}
\label{eq:rate_eq_of_n_s_p_t_in_steady_state_for_scale_free_with_facet_degree_for_d_3}
n_s(p) &= p\Bigl[ \sum_{i,j,k=1}^{\infty}{ ( {\bar{J}_{i}n_{i} \bar{J}_{j}n_{j} \bar{J}_{k}n_{k}} ) \delta_{i+j+k, s}} -3 \bar{J}_{s}n_s  \Bigr] +\delta_{1s},
\end{align}
where $n_s(p)=N_s(p,t)/N$ and $\bar{J}_{s} \equiv (s(3/2+a)-3/2)(3p+a)^{-1}$.
Thus Eq. \eqref{eq:rate_eq_of_n_s_p_t_in_steady_state_for_scale_free_with_facet_degree_for_d_3} become 
\begin{align}
\label{eq:rate_eq_of_n_s_p_t_in_steady_state_for_scale_free_with_facet_degree_2_for_d_3}
n_s(p) &= p\Bigl[ \Bigl(\frac{3/2+a}{3p+a}\Bigr)^{3}\sum_{i,j,k=1}^{\infty}{\Bigl( \Pi_{\alpha=1}^{3} {\Bigr(i_{\alpha}-\frac{3/2}{3/2+a}\Bigr)n_{i_{\alpha}}} \Bigr) \delta_{\sum_{\alpha=1}^{3}{i_{\alpha}}, s}}-3\Bigl(\frac{3/2+a}{3p+a}\Bigr)\Bigr(s-\frac{3/2}{3/2+a}\Bigr)n_{s}  \Bigr] +\delta_{1s},
\end{align}
By solving numerically this equation up to a truncated cluster size $s^{*}$, the plot of $n_s(p)$ vs $p$ for $a=0.1$ is presented in Fig.~S1.
\begin{figure}[!h]
\center
\includegraphics[width=0.6\linewidth]{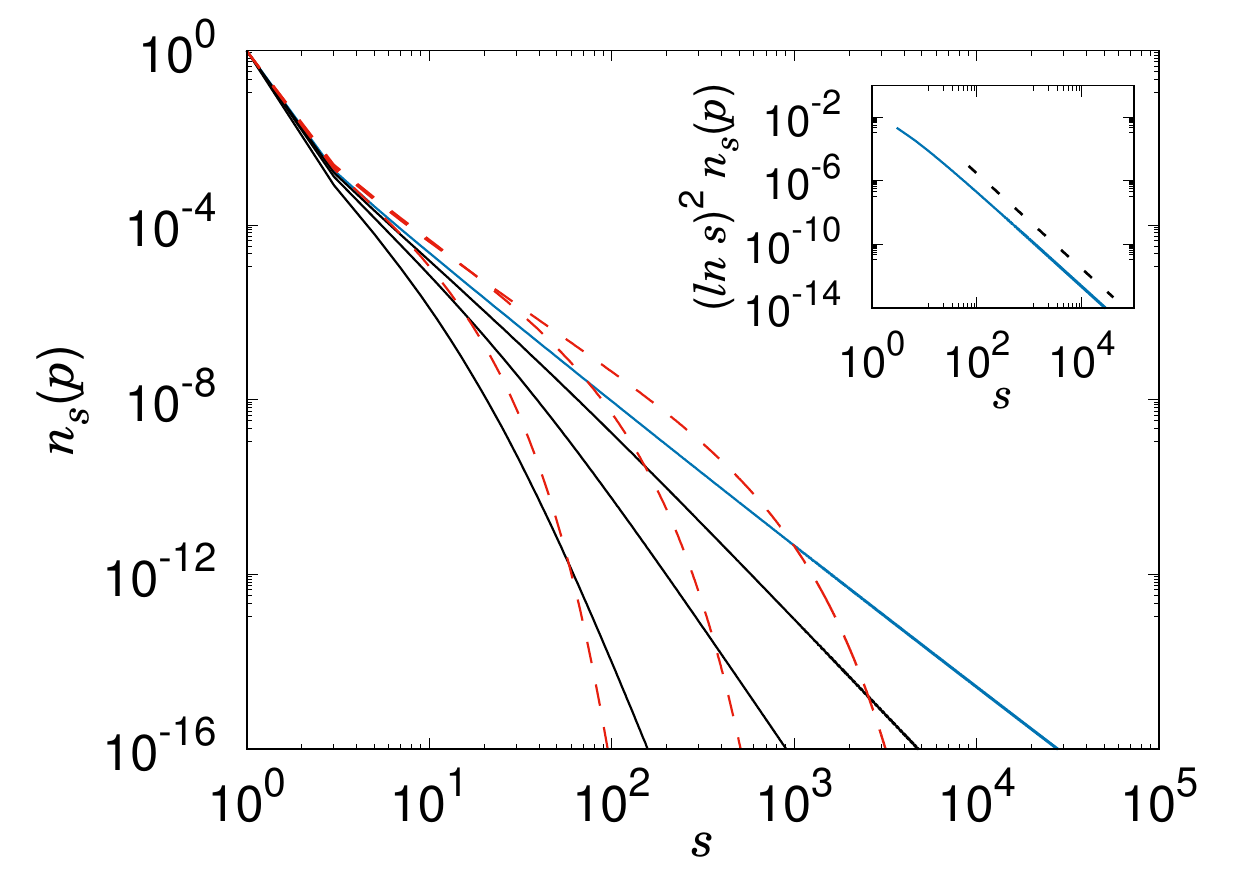}
\caption{{\bf Plot of cluster size distribution.} $n_s (p)$ versus $s$ at $p = p_c = (27-8\sqrt{11})/150 = 0.00311\cdots$ (blue solid line), for $p > p_c$ (red dashed curves), and for $p < p_c$ (black solid lines) based on numerical values obtained from the rate equation for $a=0.1$. For $p \le p_c$, $n_s (p)$ decays in a power-law manner for large $s$, indicating that the transition is infinite-order. Inset: Plot of $(\ln~s)^2 n_s (p)$ versus $s$ at $p = p_c$. The black dashed line is guideline with slope $-3$.
}
\label{fig:fig6}
\end{figure}\\

Next we introduce the two generating functions $f(x) \equiv \sum_{s=1}^{\infty}{sn_{s}x^{s}}$ and $g(x) \equiv \sum_{s=1}^{\infty}{n_{s}x^{s}}$, and then define 
\begin{align}
\label{eq:definition_of_h_for_d_3}
h(x) \equiv \frac{3/2+a}{3p+a} \Bigl( f(x) - \frac{3/2}{3/2+a}g(x) \Bigr).
\end{align}

%
%
%
%
Then $f(x)$ and $g(x)$ are written as

\begin{align}
\label{eq:The_form_of_f_for_d_3}
f(x) &= x - 3px\frac{dh(x)}{dx} + px\frac{dh^{3}(x)}{dx} = x + 3pxh'(x)(h^{2}(x) - 1), \\
\label{eq:The_form_of_n_for_d_3}
g(x) &= \int_{0}^{x}{dx' \frac{f(x')}{x'}} = x - dph(x) + ph^3(x), 
\end{align}

%
%
Then the giant cluster size $G$ and the mean cluster size $\langle s \rangle$, defined as $G = 1 - f(1)$ and $\langle s \rangle = f'(1)$, respectively, are expressed in terms of $h(1)$ and $h^\prime(1)$ as follows:
\begin{align}
\label{eq:The_form_of_giant_cluster_size_for_d_3}
G &= 1 - \frac{3p+a}{3/2+a} h(1) - \frac{3/2}{3/2+a} ( 1 - 3ph(1) + ph^3(1) ) , \\
\label{eq:The_form_of_second_order_moment_for_d_3}
\langle s \rangle &= \frac{3p+a}{3/2+a} h'(1) + \frac{3/2}{3/2+a} ( 1 - 3ph'(1) + 3ph'(1)h^{2}(1) ).
\end{align}
Furthermore, substituting Eqs. \eqref{eq:The_form_of_f_for_d_3} and \eqref{eq:The_form_of_n_for_d_3} into \eqref{eq:definition_of_h_for_d_3}, one can get
\begin{align}
\label{eq:definition_of_h_2_for_d_3}
h(x) = \frac{3/2+a}{3p+a} \Bigl( x + 3pxh'(x)(h^{2}(x) - 1) - \frac{3/2}{3/2+a} ( x -3ph(x) + ph^3(x) ) \Bigr).
\end{align}
Rearranging for $h'(z)$, Eq. \eqref{eq:definition_of_h_2_for_d_3} is rewritten as
\begin{align}
\label{eq:definition_of_hp_for_d_3}
h'(x) = \frac{ (a - (3/2)p)h(x) - ax + (3/2)ph^3(x) }{3p(3/2+a)x(h^{2}(x) - 1)}.
\end{align}
To evaluate $h(1)$ and $h^\prime(1)$, we consider Eq.~\eqref{eq:definition_of_hp_for_d_3} for two cases: in non-percolating and percolating phases. (i) In non-percolating phase, the giant cluster is absent. Thus $f(1)=1$. Using the relation $\sum_{s=1}{\Bigr[s(3/2+a) - 3/2\Bigr]N_s(t)} = 3pt + aN(t)$, we obtain $(3/2 + a)f(1) - (3/2)g(1) = 3p + a$ for large $t$. Thus, $h(1) = 1$. By applying the L’Hopital’s rule at $x=1$, one gets
\begin{align}
\label{eq:definition_of_h_when_x_1_for_d_3}
h'(1) = \frac{ (a - (3/2)p)h'(1) - a + (9/2)ph'(1) }{3p(3+2a)h'(1)},
\end{align}
leading to 
%
%
%
\begin{align}
\label{eq:solution_of_h_when_x_1_for_d_3}
h'(1) = \frac{a+3p \pm \sqrt{ a^{2} - 30ap -24a^{2}p + 9p^{2} }}{ 6p(3+2a)}.
\end{align}
The only single solution with minus sign in Eq. \eqref{eq:solution_of_h_when_x_1_for_d_3} is valid because $h'(1)$ must be zero when $p=0$. Moreover, this solution is valid for $0 \le p \le p_{c}$, where the discriminant $D(p) \equiv a^{2} - 30ap -24a^{2}p + 9p^{2}$ is non-negative value. Subsequently, the critical point $p_c$, satisfying $D(p_c) = 0$, is written as
\begin{align}
\label{eq:pc_of_growing_scale_free_simplicial_complexes_for_d_3}
p_{c}(a)= \frac{5a+4a^{2} \pm \sqrt{ 16a^{4} + 40a^{3} + 24a^{2} }}{3}.
\end{align}
Only the solution with the minus sign in Eq.~\eqref{eq:pc_of_growing_scale_free_simplicial_complexes_for_d_3} is valid to satisfy the known result $\lim_{a \to \infty}p_c (a) = 1/24$ in the limit of growing random complexes. Moreover, it is easily confirmed that the asymptotic value of $p_c(a)$ goes to $0$ for $a \to 0$. In this phase, the giant cluster size $G$ and the mean cluster size $ \langle s \rangle$ can be obtained using Eqs.~\eqref{eq:The_form_of_giant_cluster_size_for_d_3} and \eqref{eq:The_form_of_second_order_moment_for_d_3} with $h(1)$ and $h'(1)$ obtained above for $0 \le p \le p_{c}$. Numerical results are presented in Fig.~S2

(ii) In percolating phase, $f(1)$ is nonzero because the giant cluster is present. Then $h(1)$ becomes $h(1) = \int_{x=0}^{x=1} {h'(x) dx}$ and then $h'(1)$ is obtained from $h(1)$ using Eq.~\eqref{eq:definition_of_hp_for_d_3}. These values determine the giant cluster size $G = 1 - f(1)$ and $\langle s \rangle = f'(1)$ from Eqs.~\eqref{eq:The_form_of_giant_cluster_size_for_d_3} and \eqref{eq:The_form_of_second_order_moment_for_d_3}. 

In summary, the general forms of $G$ and $\langle s \rangle$ are written as
\begin{align}
\label{eq:form_of_G_in_growing_scale_free_for_d_3}
G =
\begin{cases}
0  & \textmd{for}~~~~ p < p_c , \\
\\
1 - \frac{3p+a}{3/2+a} h(1) - \frac{3/2}{3/2+a} ( 1 - 3ph(1) + ph^3(1) )  & \textmd{for}~~~~ p \geq p_c ,
\end{cases}
\end{align}
\begin{align}
\label{eq:Second_moment_form_of_G_in_growing_scale_free_for_d_3}
\langle s \rangle =
\begin{cases}
\frac{3p+a}{3/2+a} \Bigl( \frac{a+3p - \sqrt{a^{2} - 30ap -24a^{2}p + 9p^{2}}}{6p(3+2a)} + \frac{3/2}{3p+a} \Bigr)  & \textmd{for}~~~~ p < p_c , \\
\\
\frac{3p+a}{3/2+a} h'(1) + \frac{3/2}{3/2+a} ( 1 - 3ph'(1) + 3ph'(1)h^{2}(1) ) & \textmd{for}~~~~ p \geq p_c ,
\end{cases}
\end{align}
where $h(1)$ and $h'(1)$ are 
$$h(1) = \int_{x=0}^{x=1} {\frac{ (a - (3/2)p)h(x) - ax + (3/2)ph^3(x) }{3p(3/2+a)x(h^{2}(x) - 1)} dx}$$ and $$h'(1) = \frac{ (a - (3/2)p)h(1) - a + (3/2)ph^3(1) }{3p(3/2+a)(h^{2}(1) - 1)}$$ for $p \ge p_{c}$. 

We remark that 
\begin{align}
\label{eq:Second_moment_form_of_G_in_growing_scale_free_for_d_3_at_pcminus}
\langle s \rangle =
\begin{cases}
\frac{(a+3p_{c})}{3/2 + a} \Bigl( \frac{a+3p_{c}}{6p_{c}(3+2a)} + \frac{3/2}{3p_{c}+a} \Bigr)  & \textmd{as}~~~~ p \to p_c^- , \\
\\
\frac{(a+3p_{c})}{3/2 + a} \Bigl( \frac{a+3p_{c}}{3p_{c}(3+2a)} + \frac{3/2}{3p_{c}+a} \Bigr) & \textmd{as}~~~~ p \to p_c^+ ,
\end{cases}
\end{align}
Therefore there exists a discontinuity $(a+3p_{c})/[6p_{c}(3+2a)]$ for $a > 0$ between the two $\langle s \rangle$ as $p \to p_c^+$ and $p \to p_c^-$. This, together with the fact the cluster size distribution follows a power-law decay for $p < p_c$, indicates that the percolation transition is of infinite order. The numerical results are shown in Fig.~S2.

\begin{figure}[!h]
\center
\includegraphics[width=1.0\linewidth]{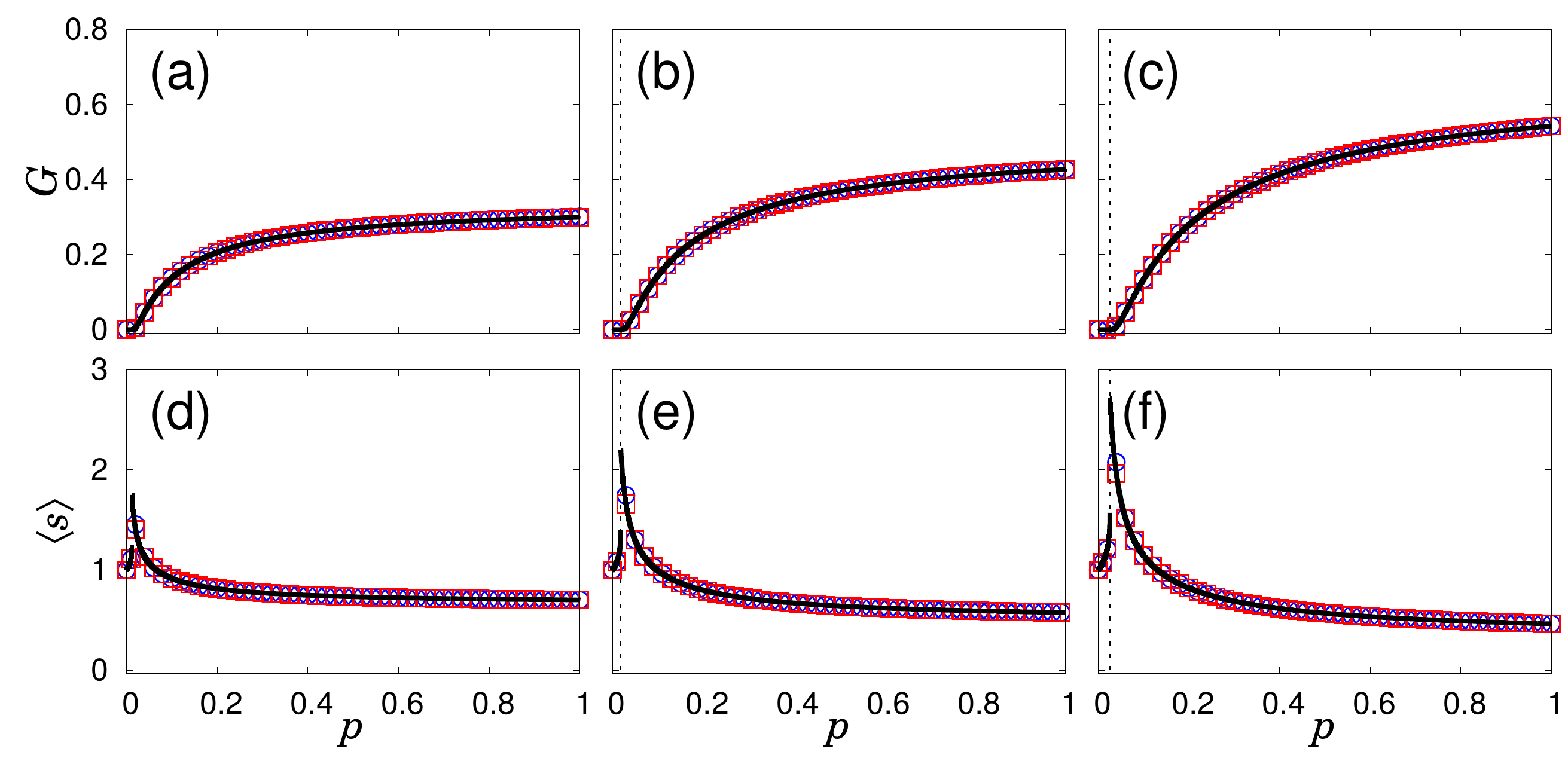}
\caption{{\bf Plots of the giant cluster size $G$ and the second order moment $\langle s \rangle$ versus $p$.} (\textbf{A} and \textbf{D}) For $a=0.5$. (\textbf{B} and \textbf{E}) For $a=1.0$. (\textbf{C} and \textbf{F}) For $a=2.0$. The blue open circle ($\textcolor{blue}{\bigcirc}$), and red open square symbols ($\textcolor{red}{\square}$) represent the Monte Carlo simulation results for the final system size $N=10^6$ and the numerical solutions of the rate equations in the steady state, respectively. Black solid curves are calculated from $f(1)$ and $f'(1)$ for $G$ and $\langle s \rangle$. The transition point, derived from analytic solution of Eq.~\eqref{eq:pc_of_growing_scale_free_simplicial_complexes_for_d_3}, is described by the dotted line for given $a$.
}
\label{fig:fig7}
\end{figure}

Assuming that $G(p)$ behaves as $\sim \exp\big(-\alpha(p-p_{c})^{-\beta}\big)$ with $\alpha > 0$ and $\beta > 0$, the exponent $\beta$ for different $a$ are obtained numerically, which are listed in Table~S1. 

\begin{table}[h!]
	\centering
	\caption{The exponent values $\beta(a)$ for different $a$.}
	\begin{tabular}{cc}
		\hline\hline
		~~~~~~$a$~~~~~~ &~~~~~~~~~~~~~~~ $\beta(a)$~~~~~~~~~~~~~~~    \\ \hline
		1/10     &  0.437(5) 		\\
		1/8      &  0.439(5) 	 	\\
		1/4      &  0.447(5)  		\\
		1/2      &  0.457(3) 		\\
		1        &  0.466(2)        \\
		2        &  0.473(1)        \\
		4        &  0.477(1)        \\
		8        &  0.480(1)        \\
		16       &  0.481(1)        \\
		32       &  0.482(1)        \\
		$\infty$ &  0.484(1)        \\ 
		\hline
	\end{tabular}
\label{table_1}
\end{table}

\subsection{Graph and facet degree distributions}

The probability $q(k_i,p,t_0,t)$ that a node $i$ introduced to the system at time $t_0$ gets a new link to existing nodes at time $t$ is proportional to $k_i(t,t_0)+a$, where $k_i(t,t_0)$ is the facet degree of node $i$ at time $t$ and $a$ is the constant. The rate equation of $q(k,p,t_0,t)$ is written as
\begin{align}
\label{eq:rate_eq_of_Fq_k_T_t_for_d_3}
q(k,p,t_0,t+1) &= p \Bigl[ \sum_{\ell=1}^{3} { \ell \binom{3}{\ell} \Bigl( \frac{k-1+a}{3pt + aN(t)}\Bigr)^{\ell}\Bigl(1-\frac{k-1+a}{3pt+aN(t)} \Bigr)^{3-\ell} q(k-1,p,t_0,t)} \nonumber \\
&+ \Bigl( 1- \frac{k + a}{3pt + aN(t)} \Bigr)^{3} q(k,p,t_0,t)\Bigr]+(1-p)q(k,p,t_0,t),
\end{align}
where the total number of nodes is $N(t)=N_0+t$. Using the facet degree distribution at $t$ defined as $P_f(k,p,t) \equiv \sum_{t_{0}=0}^{t}{{N_{0}}^{\delta_{0t_{0}}}q(k,p,t_{0},t)/N(t)}$, Eq.~\eqref{eq:rate_eq_of_Fq_k_T_t_for_d_3} is rewritten as 
\begin{eqnarray}
\label{eq:rate_eq_of_Fq_k_t_for_d_3}
N(t+1)P_f(k,p,t+1) &=& p \Bigl[ \Big(\frac{k - 1 + a}{pt/N(t) + a/3} \Big) P_f(k-1,p,t) + \Bigl( N(t) - \frac{k + a}{pt/N(t) + a/3} \Bigr) P_f(k,p,t) \Bigr] \nonumber \\
&+& (1-p)N(t) P_f(k,p,t) + q(k,p,t+1,t+1) + O\Bigl(\frac{P_f}{N(t)}\Bigr),
\end{eqnarray}
The boundary condition is given as $q(k,p,t,t)=\delta(k)$. In the limit $t\to \infty$, Eq.~\eqref{eq:rate_eq_of_Fq_k_t_for_d_3} is rewritten as
\begin{align}
\label{eq:rate_eq_of_Fq_k_t_for_d_3}
\frac{d (tP_f(k,p,t))}{d t} &= \Big(\frac{k - 1 + a}{1 + a/(3p)} \Big) P_f(k-1,p,t) - \Bigl( \frac{k + a}{1 + a/(3p)} \Bigr) P_f(k,p,t) + \delta(k).
\end{align}
Assuming that $P_f(k,p,t)$ becomes time-independent in the steady state, one easily obtains that 
\begin{align}
\label{eq:rate_eq_of_FP_k_t_in_steady_state_for_d_3}
\Bigl(1 + \frac{a}{3p} \Bigr)P_f(k,p) +  (k + a) P_f(k,p) - (k - 1 + a) P_f(k-1,p) = \Bigl(1 + \frac{a}{3p} \Bigr) \delta(k).
\end{align}
Next, using the generating function $\Phi(z,p)$ of the facet degree distribution $P_f(k,p)$ defined as $\Phi(z,p)\equiv \sum_{k=0}^{\infty} {P_f(k,p)z^{k}}$, one obtains that
\begin{align}
\label{eq:rate_eq_of_FP_k_t_using_generating_func_for_d_3}
z(1-z)\Phi' + a(1-z)\Phi + \Bigl(1 + \frac{a}{3p} \Bigr)\Phi = \Bigl( 1 + \frac{a}{3p} \Bigr).
\end{align}
\begin{figure}[!b]
\center
\includegraphics[width=1.0\linewidth]{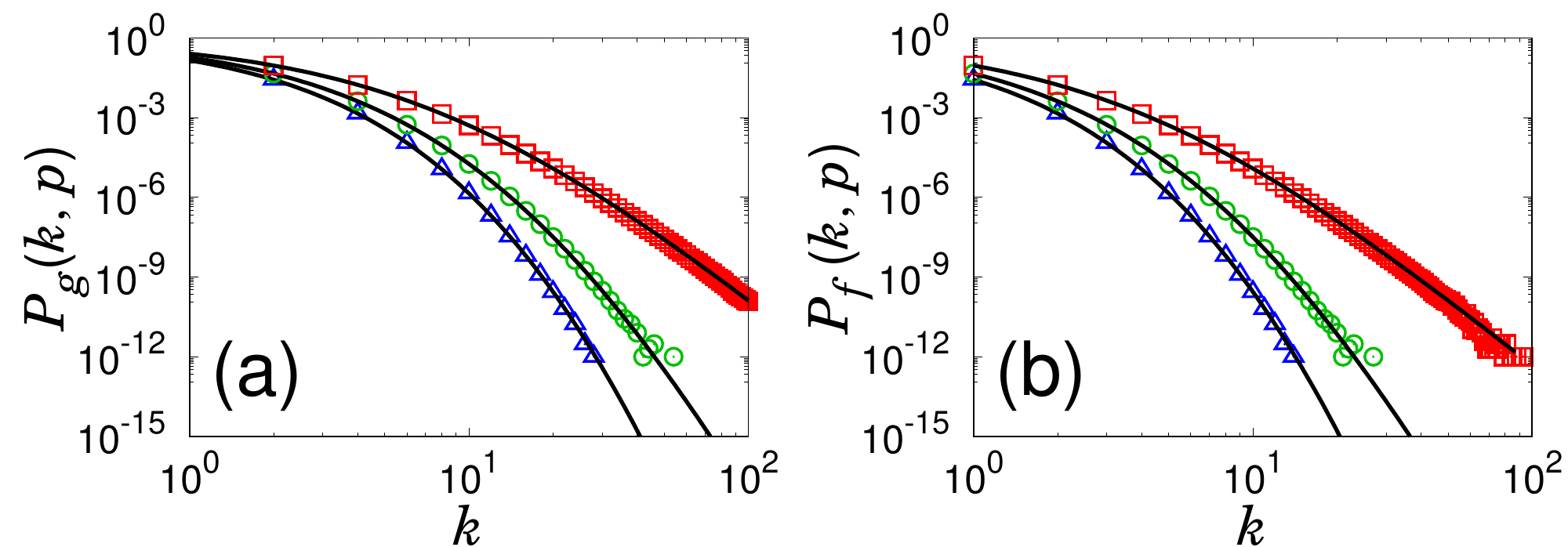}
\caption{
{\bf Plots of degree and facet degree distributions.} (\textbf{A} and \textbf{B}) Degree distribution $P_g(k,p)$ and facet degree distribution $P_f(k,p)$ versus $k$ in growing scale-free complexes. Symbols represent the simulation data of $N=10^8$ for given $p = 0.01 ~(\textcolor{blue}{\triangle})$, $p=p_c=(9-4\sqrt{5})/3=0.018576\cdots ~(\textcolor{green}{\bigcirc})$, and $p=0.05 ~(\textcolor{red}{\square})$, respectively, for given $a=1$. Solid curves are calculated from the solutions of the rate equations of the degree and facet degree distributions. Even though the theory predicts that both types of degree distributions follow power-law for large $k$; however, the numerical data for finite systems still show heavy-tailed distributions. 
}
\label{fig:fig8}
\end{figure}
The solution of \eqref{eq:rate_eq_of_FP_k_t_using_generating_func_for_d_3} around $z = 0$ is written in terms of the hypergeometric function ${}_{2}F_{1}$ as follows:
\begin{align}
\label{eq:solution_ofrate_eq_of_FP_k_t_using_generating_func_for_d_3}
\Phi(z,p) = \frac{1+a/(3p)}{1+a+a/(3p)} {}_{2}F_{1}[1,a;2+a+\frac{a}{3p};z].
\end{align}
From the relation ${}_{2}F_{1}[a,b;c;z] = \sum_{k=0}^{\infty} {\frac{1}{k!} \Bigl( \Gamma[k+a]\Gamma[k+b]\Gamma[c] \Bigr) \Bigl(\Gamma[a]\Gamma[b]\Gamma[k+c] \Bigr)^{-1} z^{k}}$, the facet degree distribution is finally written as
\begin{align}
\label{eq:Facet_degree_dist_in_GSFSC_for_d_3}
P_f(k,p) = \Bigl(1+\frac{a}{3p} \Bigr) \frac{\Gamma[1+a+a/(3p)]}{\Gamma[a]} \frac{\Gamma[k + a]}{\Gamma[2+ k + a + a/(3p)]}.
\end{align}
For $k \gg 1$, using the Striling approximation, Eq. \eqref{eq:Facet_degree_dist_in_GSFSC_for_d_3} becomes
\begin{align}
\label{eq:Facet_degree_dist_in_GSFSC_for_large_k_for_d_3}
P_f(k,p) \simeq \Bigl(1+\frac{a}{3p} \Bigr) \frac{\Gamma[1+a+a/(3p)]}{\Gamma[a]} (k + a)^{-(2+a/(3p))}.
\end{align}
Therefore, the exponent of facet degree distribution is obtained as $\lambda_f = 2 + a/(3p)$.

The graph degree distribution can be obtained similarly. Because two links are added to a selected node when a triangle is added, the graph degree distribution $P_g(k;p)$ becomes
\begin{align}
\label{eq:Degree_dist_in_GSFSC_for_d_3}
P_g(k,p) = \Bigl(1+\frac{a}{3p} \Bigr) \frac{\Gamma[1+a+a/(3p)]}{\Gamma[a]} \frac{\Gamma[k/2 + a]}{\Gamma[2+ k/2 + a + a/(3p)]}\sim k^{-[2+a/(3p)]}.
\end{align}
Therefore the exponent of the graph degree distribution is the same as the one of the facet degree distribution. That is, $\gamma_g= 2 + a/(3p)$ for large $k$, which is consistent with the one of facet degree distribution. The degree and facet degree distributions for $a=1$ are presented in Fig.~S3.	
We remark that the cluster size distribution exhibits power-law decay for $p < p_c$; however, the graph and facet degree distributions exhibit power-law decays in the entire region of $p$. 

\subsection{Homological percolation transitions}

Here, we consider a general case of the minimal model: at each time step, $d+1$ nodes are selected with the probability proportional to their facet degrees $\{k_i+a\}$. They are connected and regarded as a $d$-simplex with probability $p$. As a function of $t$, we measure the Betti number, $B_{n,d}(p,t)$, where $n$ represents the $n$-th Betti number generated by $d$-simplex. This quantity depends on the probability $p$ and time $t$.  

We find numerically that the first Betti number $B_{1,d}(p,t)$ is extensive to time $t$ (i.e., the system size $N(t)$) for $d=2,3,$ and 4. So, we express that $B_{1,d}(p,t)=b_{1,d}(p)t$ for large $t$. Then, we consider a homological percolation transition (HPT) with the normalized first Betti number $b_{1,d}(p)$ as a function of $p$. There exists a transition point $p_c^{(1,d)}$ such that the first Betti number $b_{1,d}(p)=0$ for $p \le p_c^{(1,d)}$, whereas it is finite for $p > p_c^{(1,d)}$. It is found numerically that the transition point $p_c^{(1,d)}$ is the same as that of a giant cluster. Thus, the transition point $p_c^{(1,d)}$ for $d=2$ is reduced to the analytic solution obtained in the previous section. 

Because the transition point is the same and a macroscopic homological cycle is formed in the giant cluster, one may guess that $b_{1,d}(p)$ is somewhat related to the size of the giant cluster. Thus, we try to fit the normalized first Betti number to the same form of the giant cluster, i.e., the essential singular form near the transition point as  
\begin{align}
b_{1,d}(p) \sim e^{-\alpha_{1,d}(p-p_c^{(1,d)})^{-\beta_{1,d}}}.
\end{align}


To test this functional form, we perform numerical analysis for $d=2$ as follows: First, we measure the occurrence probability $P(B_1;p,t)$ that a sample contains at least one first Betti number in the largest cluster for different $p$ and $t$. This quantity is similar to the probability that a sample contains a spanning cluster for a given occupation probability $p$ in ordinary percolation. We examine the slope $dP(B_1;p,t)/dp$ for given $t$, and then find $p^*(t)$ at which the slope becomes maximum (see Fig.~S4A). This is regarded as the transition point of the HPT of the first Betti number for given system size $t$. Next, using the finite size scaling idea, we determine the transition point $p_{c1}$ in the thermodynamic limit $t\to \infty$. That is, $p_{c1}$ satisfies the power-law relation $p^*(t)-p_{c1}\sim t^{-1/\bar{\nu}^*}$ (see Fig.~S4B). Next, using the estimated $p_{c1}$, we confirm that the first Betti number behaves in the essential singular form near the transition point, i.e. $b_1 \sim \text{exp} (-\alpha_1 (p-p_{c1})^{-\beta_1})$ (Fig.~S5A)). Using the similar method, we find that the second Betti number also follows an essential singular form as shown in Fig.~S5B. We also obtain the transition point and the critical exponents $\beta$ and $1/\bar{\nu}$ for different attractiveness $a$, which are listed in Table~S2.

\begin{table}[!h]\centering
\caption{Transition point $p_c$ and the critical exponents $\beta$ and $1/\bar{\nu}$ of the first and second Betti numbers for different $a$.}
\begin{tabular}{c|cc||c|cc||c|cc}
\hline
$a=0.1$ & $b_1$ & $b_2$ & $a=0.25$ & $b_1$ & $b_2$ & $a=0.5$ & $b_1$ & $b_2$ \\
\hline
$p_c$ & 0.0031 & 0.053(3) & $p_c$ & 0.0070 & 0.136(1) & $p_c$ & 0.0120 & 0.252(1) \\
$\beta$ & 0.434(1) & 0.95(3) & $\beta$ & 0.476(2) & 1.34(1) & $\beta$ & 0.493(2) & 2.1(1) \\
$ 1/\bar{\nu}$ & 0.124(2) & 0.058(2) & $ 1/\bar{\nu}$ & 0.123(2) & 0.048(2) & $ 1/\bar{\nu}$ & 0.120(2) & 0.032(2) \\
\hline

\end{tabular}
\label{table2}
\end{table}

\begin{figure}[!t]
\center
\includegraphics[width=1.0\linewidth]{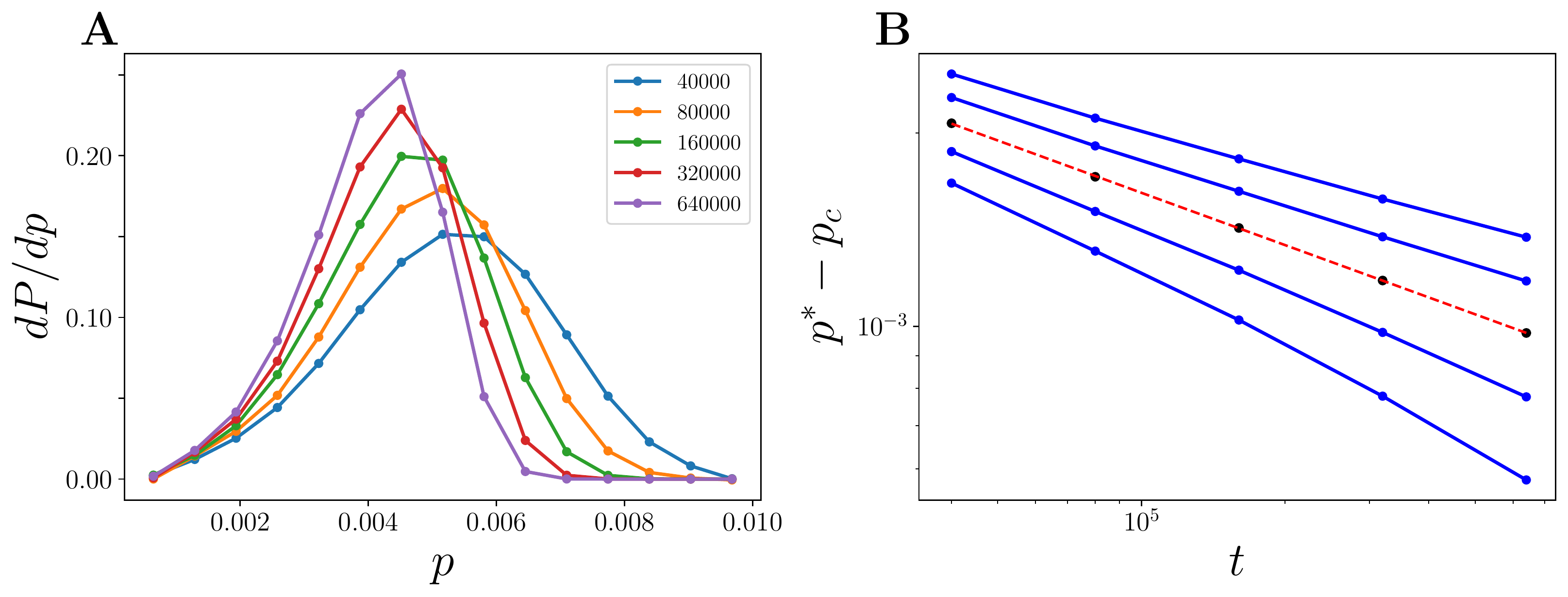}
\caption{{\bf Determination of the transition point $p_c$} In (\textbf{A}), plot of $dP/dp$ versus $p$, and determine peak positions for each $t$. Legend indicates different $t$ values. In (\textbf{B}), plot of  $p^*(t)-p_c$ versus $t$ for different trials of $p_c$, and find $p_c$ that exhibits power-law decay. The slope of the power-law behavior determines $1/\bar{\nu}^*$.
}
\label{fig:fig9}
\end{figure}

\begin{figure}[!h]
\center
\includegraphics[width=1.0\linewidth]{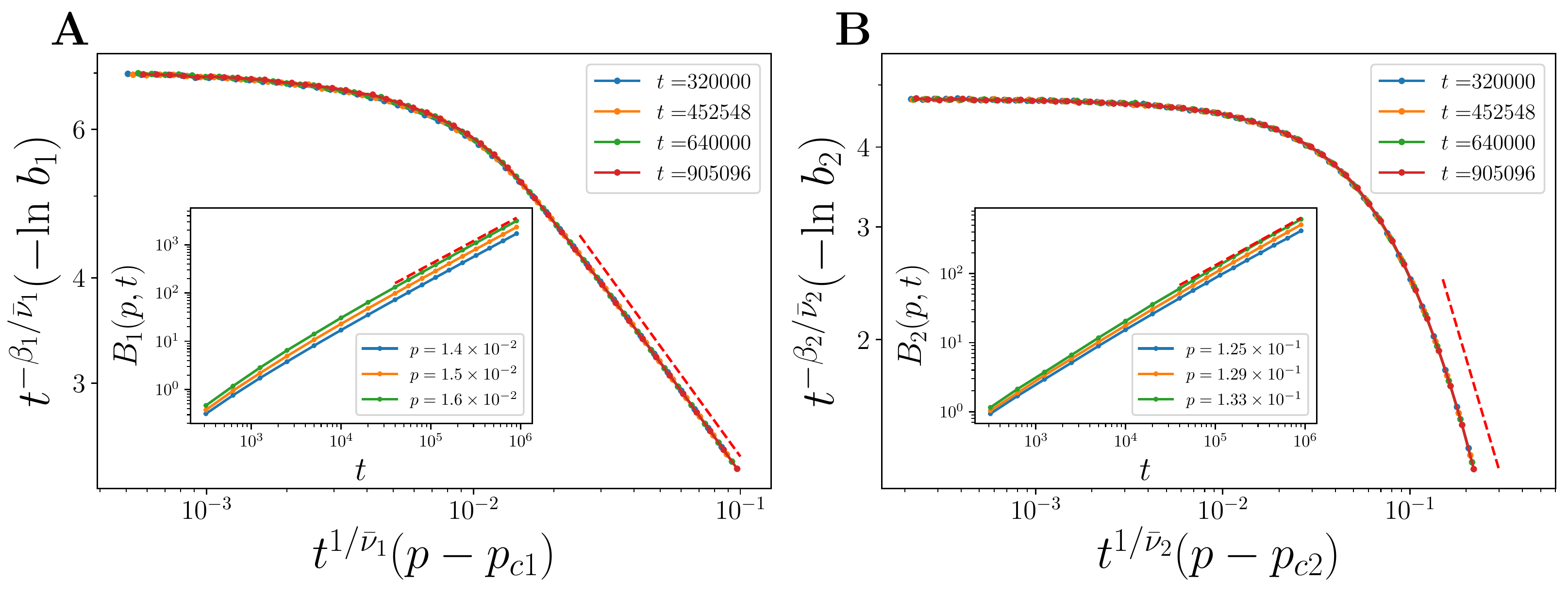}
\caption{(Color online)
{\bf Scaling plots of the normalized first and second Betti numbers.} To test the essential singular behaviors of $b_1(p)$ and $b_2(p)$, we plot $t^{-\beta_k/\bar{\nu_k}}(-\ln~b_k)$ versus $t^{1/\bar{\nu_k}}(p-p_{ck})$, where $k=1$ for (\textbf{A}) and $k=2$ for (\textbf{B}). Dashed lines are guidelines with slopes $-\beta_1$ for (A) and $-\beta_2$ for (B) given in Table~S2. Insets: plots of $B_1(p,t)$ (A) and $B_2(p,t)$ (B) versus $t$ for different $p$ values in double-logarithmic scales. They show power-law behaviors $B_1 \sim t$ and $B_2 \sim t^{0.70}$.  
}
\label{fig:fig10}
\end{figure}

\subsection{Model for protein interaction complex.}
The protein-interaction SC (PISC) model has two processes: duplication and divergence processes. The PISC model evolves in the following way: 
i) At each time step, a new vertex (denoted as $B$) is added as a mutant  of an existing vertex (denoted as $A$). Each facet of vertex $A$ is duplicated with probability $1-\alpha$ and then vertex $A$ is replaced by vertex $B$. ii) Two distinct vertices are randomly selected and make a triangle with vertex $B$. This process occurs with probability $\beta$. Processes i) and ii) mimic duplication and mutation, and divergence processes, respectively.

\newpage
\section{The relation between the number of nodes and dates }

\begin{table}[!h]\centering
\caption{Timeline of the coauthorship data in the interval of six months.}
\begin{tabular}{c|c|c||c|c|c}
\hline
time ($t$) & month/year & number of authors ($N_0$)&
time ($t$) & month/year & number of authors ($N_0$)\\
\hline
0&06/98&4&120&06/08&6572\\
6&12/98&6&126&12/08&7303\\
12&06/99&14&132&06/09&8125\\
18&12/99&47&138&12/09&8975\\
24&06/00&98&144&06/10&9807\\
30&12/00&165&150&12/10&10760\\
36&06/01&242&156&06/11&11906\\
42&12/01&368&162&12/11&12953\\
48&06/02&521&168&06/12&14266\\
54&12/02&714&174&12/12&15513\\
60&06/03&916&180&06/13&16848\\
66&12/03&1152&186&12/13&18334\\
72&06/04&1616&192&06/14&19821\\
78&12/04&2071&198&12/14&21436\\
84&06/05&2516&204&06/15&23036\\
90&12/05&2982&210&12/15&24787\\
96&06/06&3653&216&06/16&26564\\
102&12/06&4389&222&12/16&28451\\
108&06/07&5134&228&06/17&30306\\
114&12/07&5809&234&12/17&32016\\

\hline

\end{tabular}
\label{table3}
\end{table}